\documentclass[11pt]{article}
\usepackage{amssymb,latexsym,amsmath, amsthm}
\usepackage{amsfonts, graphics,color}
\usepackage{mathtools}
\usepackage{hyperref} 
\usepackage{breakurl} 
\usepackage[american]{babel}
\usepackage{csquotes}
\usepackage{enumitem}
\usepackage{dsfont}
\usepackage[utf8]{inputenc}

\usepackage[top=1in, bottom=1.25in, left=1.25in, right=1.25in]{geometry}

\usepackage{subcaption} 
\usepackage{float}

\def\R{\mathbb R}

\let\originalleft\left
\let\originalright\right
\renewcommand{\left}{\mathopen{}\mathclose\bgroup\originalleft}
\renewcommand{\right}{\aftergroup\egroup\originalright}
\newcommand{\ft}[0]{\footnotesize}

\newcommand{\EVN}{EVStabilityNet}

\usepackage{xcolor}
\usepackage[normalem]{ulem} 
\def\bcr{\begin{color}{red}}
\def\bcb{\begin{color}{blue}}
\def\bcc{\begin{color}{violet}}
\definecolor{darkgreen}{RGB}{0,150,0}
\def\bcg{\begin{color}{darkgreen}}
\def\ec{\end{color}}
\def\be{\begin{equation}}
\def\ee{\end{equation}}

\hypersetup{pdftitle={GS-EVN}}

\numberwithin{equation}{section}

\title{\EVN: Predicting the Stability of Star Clusters\\ in General Relativity}

\author{Christopher~Straub  \& Sebastian Wolfschmidt \vspace{0.4cm}   \\ 
  Department of Mathematics, University of Bayreuth, Germany}

\begin{document}

\maketitle

\begin{abstract}
We present a deep neural network which predicts the stability of isotropic steady states of the asymptotically flat, spherically symmetric Einstein-Vlasov system in Schwarzschild coordinates.
The network takes as input the energy profile and the redshift of the steady state.  
Its architecture consists of a U-Net with a dense bridge. 
The network was trained on more than ten thousand steady states using an active learning scheme and has high accuracy on test data.
As first applications, we analyze the validity of physical hypotheses regarding the stability of the steady states. 
\end{abstract}

\vspace{-.5cm}

\tableofcontents

\section{Introduction}

The Einstein-Vlasov system describes the dynamics of self-gravitating, collisionless matter by coupling Albert Einstein's general relativity with the Vlasov (or collisionless Boltzmann) equation. This system holds strong significance in astrophysics, from modeling the evolution of star clusters and galaxy clusters to understanding the formation of black holes~\cite{An2011, AnKuRe10, AnKuRe11, Ehlers73, IT68, Rein95, Gue23}.
An ensemble of particles or gas---interacting solely via their self-generated gravitation---is treated as a continuous distribution of mass represented by a phase-space density function~$f$. We are only interested in the asymptotically flat case with a vanishing cosmological constant. The speed of light and the gravitational constant are normalized to unity. We always assume spherical symmetry, i.e., we consider ball-shaped star clusters. 
We state the system in this setting in Section~\ref{ssc:ev}.
For an overview of what is known about the Einstein-Vlasov system, we refer the interested reader to~\cite{An2011,Rein95, Gue23} and the references therein.

The Einstein-Vlasov system possesses a plethora of steady state solutions.
Such a solution corresponds to a configuration with constant-in-time density.
Hence, steady states are common models for star clusters which have settled in an equilibrium state.
In this work we investigate isotropic steady states of the Einstein-Vlasov system.
This means that the phase space density of the steady state depends only on the particle energy.
The density of particles with a certain energy is determined by the prescribed {\em energy profile function} $\Phi\colon[0,1[\to[0,\infty[$; see Section~\ref{ssc:stst} for a concrete description.
A physically motivated~\cite{Ip1969_1, Ze1971} assumption is that~$\Phi$ is strictly increasing. This results in the concentration of ever more energetic particles to be decreasing within the equilibrium configuration.
Another parameter of the steady state is the redshift factor $\kappa>0$.
Loosely speaking, large values of~$\kappa$ correspond to highly relativistic configurations, while choosing smaller values of~$\kappa$ lead to less relativistic steady states.
Under suitable assumptions on~$\Phi$, any combination of~$\Phi$ and $\kappa>0$ yields a compactly supported steady state of the Einstein-Vlasov system.
We review the steady state construction in more detail in Section~\ref{ssc:stst}.

Stability theory analyzes the response of steady states to small perturbations.
For instance, a star cluster in equilibrium might be perturbed by the gravitational force of other star clusters passing by.
Loosely speaking, a {\em stable} steady state remains close to the equilibrium configuration after perturbation, while an {\em unstable} steady state evolves away from the equilibrium.
In the present setting, slightly perturbing an unstable steady state either carries (parts of) the configuration to a new equilibrium or leads to the collapse into a black hole~\cite{Praktikum20}.
Obviously, we cannot expect to meet in nature an unstable steady state~\cite[Ch.~5]{BiTr}.
We must therefore first determine the stability of a steady state that is to be used as a model in reality.

Unfortunately, analyzing the stability of the aforementioned steady states turns out to be very challenging.
In fact, neither the stability nor the instability of any such steady state has yet been mathematically proven.
We refer to~\cite{Re23} for a recent review of the mathematical results towards this question.
Let us explicitly mention here the following two results proven on the linearized level, i.e., in a simplified yet physically well motivated setting: 
For a fixed energy profile function~$\Phi$ (satisfying suitable regularity assumptions), any not too relativistic steady state ($0<\kappa\ll1$) is stable~\cite{HaRe2014}, while any highly relativistic steady state ($\kappa\gg1$) is unstable~\cite{HaLiRe2020}. 
It is conjectured that this also holds for the actual non-linear stability behavior.
Furthermore, in the astrophysics literature, some hypotheses regarding the precise location of the onset of instability, i.e., the $\kappa$-value where the steady states change from being stable to being unstable, have been developed.

Since the 1960's, the stability of steady states of the Einstein-Vlasov system has also been analyzed numerically.
We refer to~\cite[Sc.~1.2]{GueStRe21} and~\cite[Sc.~8]{Re23} for an overview over past numerical investigations.
The motivation is that a thorough understanding of the stability of steady states, derived numerically, ought to be useful for proving the stability properties mathematically.  
For instance, a recent numerical investigation~\cite{GueStRe21} falsified a long-standing stability hypothesis from the astrophysics literature.
In numerical stability investigations, one first computes a steady state, perturbs it slightly, and then evolves the resulting configuration over time using a suitable simulation of the Einstein-Vlasov system.
We describe our explicit numerical algorithm in Section~\ref{ssc:numeric}.
Although the numerical simulations have become very accurate and reliable over the years, they are still computationally expensive.
Determining the stability of a large amount of steady states, e.g., to test the validity of a physical hypothesis in depth, is usually not possible due to limited computational resources. 

For this reason we have developed a deep neural network which can quickly and accurately predict the stability of steady states: the {\EVN}. 
The architecture of the \EVN\ consists of a U-Net with a dense bridge. These two parts mostly work in parallel and are unified only at late layers of the network.
This architecture is motivated by mathematical aspects of the problem.
We discuss the architecture in more detail in Section~\ref{ssc:architect}. 

The \EVN\ was trained on over $10^4$ steady states, which were labeled as \enquote{stable} or \enquote{unstable} using the numerical simulations outlined above.
In order to select the training data, we employed an active learning scheme~\cite{KuGu20,Se09}, i.e., after starting with a randomly chosen set of training data, we added more training data iteratively, see Section~\ref{ssc:traindata}.

In Section~\ref{ssc:performance} we analyze the speed and accuracy of the \EVN.
On randomly selected test data, the \EVN\ reaches an accuracy of $99\%$.
We also test the accuracy on isotropic polytropic steady states as well as on the steady states from~\cite[Fig.~4]{GueStRe21}.
The former are particularly natural from a mathematics point of view, whereas the latter exhibit highly diverse stability properties.
The \EVN\ completes all tests with convincing accuracy.

We have made the \EVN\ publicly available so that it can be tested and applied by everyone. The download link and some notes on how to use it are provided in Section~\ref{ssc:howtouse}. 

In Section~\ref{sc:applications}, we apply the \EVN\ to test the validity of several hypotheses from the astrophysics literature regarding the stability of steady states.
All these hypotheses are about linking (in)stability to a criterion that can be checked by examining specific quantities of the steady state, such as its total mass or redshift.
To investigate the validity of such a hypothesis, we compare its statements about the stability of steady states to the predictions of the \EVN. Due to the \EVN's low computational costs, we can perform this comparison on larger data sets than would otherwise be possible.
If the stability predictions of the hypothesis and the \EVN\ consistently agree, it provides a strong indication of the hypothesis's validity.
Otherwise, the \EVN\ provides promising examples of steady states for which the hypothesis does not hold. These examples can then be double-checked with the conventional particle-in-cell method.  

The first hypothesis we examine in this way is the so-called {\em weak binding energy hypothesis}. It states that if the redshift~$\kappa$ is increased for steady states with a fixed energy profile function~$\Phi$, the steady states are stable at least up to the first maximum of the binding energy $E_b$, see~\eqref{eq:Eb} for a definition of this quantity.
This hypothesis has been tested and confirmed repeatedly in the literature~\cite{AnRe2006, Praktikum20, GueStRe21,Ip1969_2,Ip1980, RaShTe1989_1,  ShTe1985_1, ShTe1985_2}. We can confirm it again here after testing it on $2000$ randomly generated energy profile functions.
The second hypothesis, the {\em strong binding energy criterion}, is more rigid than the first, claiming that the stability changes precisely at the first maximum of~$E_b$. As noted above, this hypothesis has been proven to be not valid for general steady states in~\cite{GueStRe21}. We come to the same conclusion here. It is notable that the hypothesis, nonetheless, seems to hold true for a significant proportion of the randomly selected steady states we tested it on.
The third hypothesis originates from~\cite{AnRe2006,Praktikum20,ShTe1985_2} and states that every steady state with negative binding energy has to be unstable. The predictions of the \EVN\ are in agreement with this hypothesis on a large number of randomly generated steady states.
Lastly, we outline a future application concerning the presence of oscillating and damped star clusters at the end of Section~\ref{sc:applications}.

\vspace*{.5cm}
\noindent
{\bf Acknowledgments.}
We are grateful to Gerhard Rein for his guidance and for the freedom he granted us to conduct independent research, which led to this paper.

\section{Mathematical and numerical background}

\subsection{The Einstein-Vlasov system}\label{ssc:ev}

We describe the evolution of a star cluster by its phase-space density function~$f$.
Due to the assumption of spherically symmetry, $f=f(t,r,w,L)$ can be written as a function of the time variable $t\in\R$, the radial space variable $r\geq0$, the radial momentum variable $w\in\R$, and the variable $L\geq0$ describing the squared modulus of the angular momentum.
For a fixed time~$t$, the mass of all particles with phase-space coordinates $(r,w,L)$ lying in a given set $A\subset\R^3$ equals the value of the integral $$\int_A \sqrt{1+w^2+\frac L{r^2}}\, f(t,r,w,L)\,d(r,w,L).$$
The mass density and pressure associated to~$f$ are given by
\begin{align}
	\rho_f(t,r)&=\frac\pi{r^2}\int_0^\infty\int_\R\varepsilon\,f(t,r,w,L)\,dw\,dL,\label{eq:rho}\\
	p_f(t,r)&=\frac\pi{r^2}\int_0^\infty\int_\R\frac{w^2}\varepsilon\,f(t,r,w,L)\,dw\,dL,\label{eq:p}
\end{align}
respectively, where we introduce the abbreviation
\begin{equation}
	\varepsilon\coloneqq\sqrt{1+w^2+\frac L{r^2}}.
\end{equation}
We consider Schwarzschild coordinates
\begin{equation}\label{eq_intro:schwarzschild_lineelement} 
	ds^2 = - e^{2\mu(t,r)} dt^2 + e^{2\lambda(t,r)} dr^2 + r^2(d\theta^2 + \sin^2(\theta) d\psi^2),
\end{equation} 
in which the metric parameterizing the geometry of spacetime is determined by the two functions $\mu_f=\mu_f(t,r)$ and $\lambda_f=\lambda_f(t,r)$ given by~$\rho_f$ and~$p_f$ via the differential equations
\begin{equation}\label{eq:mulambda}
	e^{-2\lambda_f}(2r \partial_r \lambda_f-1)+1=8\pi r^2\rho_f,\qquad e^{-2\lambda_f}(2r \partial_r \mu_f+1)-1=8\pi r^2p_f.
\end{equation} 
Equations~\eqref{eq:mulambda} correspond to Einstein's equations in Schwarzschild coordinates.
The assumption that we consider an isolated system with a regular center (at $r=0$)  is implemented mathematically by incorporating the following boundary conditions on~$\mu_f$ and~$\lambda_f$:
\begin{equation}
	\lim_{r\to\infty}\lambda_f(t,r)=\lim_{r\to\infty}\mu_f(t,r)=0=\lambda_f(t,0).
\end{equation}
We refer to~\cite{An2011,Rein95} for more background, but state here the following nice feature of Schwarzschild coordinates:
The variable~$t$ can indeed be interpreted as the proper time of an observer looking onto the configuration from spatial infinity.
The reason we choose these coordinates is because they simplify Einstein's equations sufficiently and because they are the most commonly used in the literature.
However, past numerical investigations suggest that choosing other coordinates does not affect the stability properties which we will analyze later~\cite{Praktikum20,GueStRe21}.
The metric coefficients $\mu_f$ and~$\lambda_f$ determine the evolution of~$f$ via the partial differential equation 
\begin{equation}\label{eq:Vlasov}
	\partial_tf+e^{\mu_f-\lambda_f}\frac w\varepsilon\,\partial_rf-\left(w\,\partial_t \lambda_f+\varepsilon\,e^{\mu_f-\lambda_f}\mu_f'-\frac L{r^3\varepsilon}\,e^{\mu_f-\lambda_f}\right)\,\partial_wf=0.
\end{equation}
This equation is known as the {\em Vlasov equation}.
Accordingly, the system~\eqref{eq:rho}--\eqref{eq:Vlasov} is the {\em asymptotically flat Einstein-Vlasov system in Schwarzschild coordinates}.
For brevity, we shall refer to it simply as the {\em Einstein-Vlasov system} throughout this article.
It is a closed system for the phase-space density function~$f$.
A derivation of the system in the above form can be found in~\cite{Rein95}. 
The existence theory is reviewed in~\cite{An2011}.

\subsection{Steady states}\label{ssc:stst}

A steady state is a time-independent solution $f_0=f_0(r,w,L)$ of the Einstein-Vlasov system~\eqref{eq:rho}--\eqref{eq:Vlasov}.
As motivated earlier, we seek such steady states by assuming that they only depend on the particle energy. 
More precisely,
\begin{equation}\label{eq:f0}
	f_0(r,w,L)=\Phi\left(1-\frac{E(r,w,L)}{E_0}\right),
\end{equation}
where the energy~$E=E(r,w,L)$ of a particle with phase-space coordinates $(r,w,L)$ is given by
\begin{equation}
	E(r,w,L)=e^{\mu_{f_0}(r)}\varepsilon,
\end{equation}
and $E_0\in\,]0,1[$ is the {\em cut-off energy} which we determine below.
For the {\em energy profile function} \mbox{$\Phi\colon[0,1[\to[0,\infty[$}, we make the following assumptions: 
\begin{enumerate}[label=($\Phi\arabic*$)]
	\item\label{it:assphi1} $\Phi$ is continuous and increasing with $\Phi(0)=0$.
	\item\label{it:assphi2} There exist constants $\eta_0>0$ and $0< k\leq2$ such that
	\begin{equation}\label{eq:growth_condition}
		\Phi(\eta)=\eta^k,\qquad0<\eta<\eta_0.
	\end{equation}
\end{enumerate}
The assumption~\ref{it:assphi2} looks rather restrictive at first glance, but we discuss it below. 
We always extend~$\Phi$ continuously onto~$]-\infty,1[$ by setting $\Phi(\eta)\coloneqq0$ for $\eta<0$, which implies $f_0(r,w,L)=0$ if $E(r,w,L)\geq E_0$.

Making the ansatz~\eqref{eq:f0} reduces the Einstein-Vlasov system for~$f_0$ to a scalar integro-differential equation for the auxiliary quantity~$ y \coloneqq \ln(E_0) - \mu_{f_0}$, cf.~\cite[Sc.~1]{ReRe93}.
It is proven in~\cite{RaRe2013} that for any initial value $y(0)=\kappa >0$ and any energy profile function~$\Phi$ satisfying~\ref{it:assphi1} and~\ref{it:assphi2} with $0<k<\frac32$, the integro-differential equation has a unique solution leading to a steady state~$f_0$ with cut-off energy $E_0$ determined by $E_0 =\exp(\lim_{r\to\infty}y(r))$. 
This steady state is compactly supported and has finite (ADM-)mass, i.e., $\{f_0>0\}$ is bounded and $M\coloneqq\int \varepsilon f_0 <\infty $, which shows that~$f_0$ indeed serves as a realistic model for a star cluster in equilibrium.
Numerical simulations show that the same properties also hold for $\frac32\leq k\leq2$, which is why we include these cases here too.

By the scaling law from~\cite[Sc.~2.3]{GueStRe21}, multiplying the energy profile function~$\Phi$ by a positive factor results in a steady state with the same stability behavior. This means that, after rescaling, our investigation also covers energy profile functions satisfying
\begin{equation}\label{eq:growth_condition_c}
	\Phi(\eta)=c\,\eta^k,\qquad0<\eta\ll1,
\end{equation}   
for some $c>0$ and~$k$ as above.
All steady states satisfying~\eqref{eq:growth_condition_c} constitute a sufficient extensive and diverse class for the numerical analysis, although mathematically, even more general energy profile functions could be considered~\cite{RaRe2013}.

In conclusion, we obtain for every energy profile $\Phi$ a family of stationary solutions which we denote as $(f_\kappa)_{\kappa>0}$. We only consider $\kappa<1$ since we have never encountered a steady state which is stable for $\kappa>0.8$.

A physically important quantity associated to a steady state~$f_0$ is its (fractional) {\em binding energy}
\begin{equation}\label{eq:Eb}
	E_b=\frac{N-M}M,
\end{equation}
where $N=\int e^{\lambda_{f_0}}f_0$ is the particle number of~$f_0$ and~$M$ is its mass.

\subsection{Numerical stability analysis}\label{ssc:numeric}

In order to determine the stability of a steady state~$f_0$, we analyze the solution~$f$ of the Einstein-Vlasov system launched by the initial distribution $\mathring f\coloneqq\alpha f_0$.
One should think of~$\mathring f$ as a perturbation of the steady state~$f_0$. 
The strength of the perturbation is determined by the difference of the {\em amplitude}~$\alpha$ and~$1$. 
As observed in~\cite{AnRe2006, Praktikum20}, choosing $\alpha>1$ in the case of an unstable steady state~$f_0$ leads to the collapse of the matter and the formation of a black hole. One manifestation of this behavior is that $e^{\mu_f(t,0)}$ gets close to~$0$.
In the case of a stable steady state, any slight perturbation, i.e., $|1-\alpha|\ll1$, results in the solution~$f$ remaining close to~$f_0$ for all times.

We evolve the initial condition~$\mathring f$ numerically using a {\em particle-in-cell scheme}. 
This is the state-of-the-art method for simulating the Einstein-Vlasov system and is also the most commonly used in the literature~\cite{AmAnRi21,AmAnRi23,AnRe2006,Praktikum20,GueStRe21,ReReSch}. 
The basic idea is to split the $(r,w,L)$-support of~$\mathring f$ into a finite number of cells.
In the center of each cell we place a numerical particle representing the contribution of its cell.
These particles are propagated according to the characteristic system associated to the Vlasov equation~\eqref{eq:Vlasov}; the arising ODE is solved using the fourth-order Runge-Kutta method.
Based on the new positions of the numerical particles, the metric coefficients are updated after each time step.
We refer the interested reader to the literature cited above for more details on the particle-in-cell scheme. 
The code realizing this numerical method is written in~\texttt{C++}. Based on~\cite{KoRaRe2013}, it is parallelized using the Pthreads API.
For the choice of all numerical parameters we have followed~\cite{GueStRe21}. 
For instance, we use roughly $7\cdot10^7\,\kappa^{\frac 32}$ numerical particles to represent the distribution function. The $\kappa$-dependency is included here in order to handle the more peaked metric coefficients occurring at more relativistic configurations.
As described in~\cite[Sc.~3]{GueStRe21}, this method simulates the Einstein-Vlasov system with high accuracy.

The amplitude $\alpha$, which determines the strength of the perturbation, is fixed at \mbox{$\alpha = 1.0001$}. In order to automatically detect a stable or unstable steady state, we distinguish between two situations:  If $e^{\mu_{f}(t,0)}$ eventually converges to zero due to the perturbation, in which case we stop the simulation, we define the steady state as unstable. Otherwise, $e^{\mu_{f}(t,0)}$ stays roughly at $e^{\mu_{f}(0,0)}$ until $ t=500 M$, and we deem the steady state stable unless we observe unusual behavior after manually inspecting $e^{\mu_{f}(t,0)}$. Note that this dichotomy has been found in all the numerical studies mentioned earlier. 

\section{The \EVN}\label{sc:net}

\subsection{Architecture}\label{ssc:architect}

\begin{figure}[h]
	\centering
	\includegraphics[width=1\textwidth]{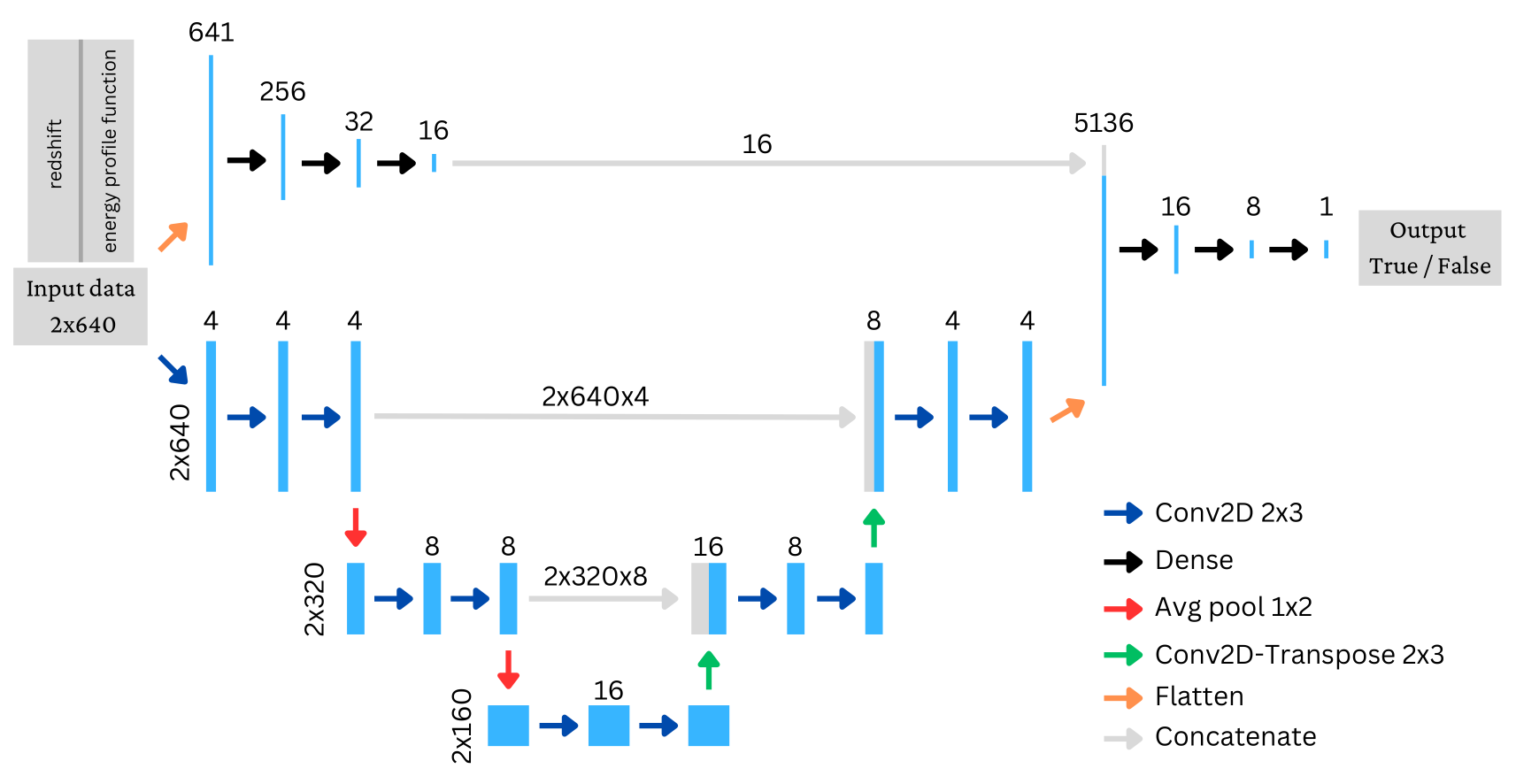}
	\caption{Architecture of the \EVN.}
	\label{img:architecture}
\end{figure}

The neural network architecture for predicting the stability of steady states of the Einstein-Vlasov system is based on a U-Net~\cite{RoFiBr15} with a dense bridge bypassing the U-Net. 
The architecture is shown in detail in Figure~\ref{img:architecture}. 

U-Nets have their origins in image segmentation tasks~\cite{RoFiBr15} and are commonly used in that field.
Their fully convolutional architecture enables them to effectively capture local features within the input data.
In the present context, this could, e.g., be the slope of the energy profile function~$\Phi$. 
The U-Net consists of an encoder and a decoder, both using convolutional and pooling layers.
To avert loss of information in the encoding stage, skip connections transfer the information from the encoding path to the decoding path.  

In addition to the fully convolutional U-Net, we integrate a dense bridge into the \EVN.
This improves the network's capacity for capturing global features in the input data.
This is useful in our context because global information about the steady state, such as its total mass or particle number, is connected to its stability in certain cases~\cite{Praktikum20,GueStRe21}.

As an aside, we have alternatively tried using a pure dense network, a pure U-Net, and different convolutional networks instead of the U-Net, but this has led to worse results at similar training costs.

The network takes as input a steady state as a two-dimensional tensor of size $2\times640$, i.e., two channels of length $640$.
The first channel contains the redshift value~$\kappa>0$ repeated $640$ times; this is done to ensure access to the value of~$\kappa$ in the convolutional layers.
The second input channel stores the values of the energy profile function~$\Phi$, discretized with a step size of $\Delta\eta=0.001$. 
We further set the values of~$\Phi$ outside $[0,1-e^{-\kappa}]$ to~$0$ as these values have no significance in the computation of the steady state, cf.\ Section~\ref{ssc:stst}. 
Because we always choose $\kappa<1$, this also shows that we generally only need the values of~$\Phi$ on \mbox{$[0,1-e^{-1}]\subset[0,0.639]$}, leading to the input length of $640$.
An example of the input data is given in Table~\ref{table:input}. 
For the dense path, we flatten the input tensor and delete the copies of the value of~$\kappa$.

\begin{table}[h]
	\centering
	\begin{tabular}{c||c|c|c|c|c|c|c|c|c|c}
		Index Nr.& $0$ & $1$ & $2$ & $\cdots$ & $180$ & $181$ & $182$ & $183$ & $\cdots$ & $639$\\
		\hline\hline
		$\kappa$ & $0.2$ &$ 0.2$ &  $0.2$ & $\cdots$ & $0.2$ & $0.2$ & $0.2$ & $0.2$ & $\cdots$ &$ 0.2$ \\
		\hline
		$\Phi(\eta)$ & $0$ & $0.001$ &  $0.002$ & $\cdots$ & $0.18$ & $0.181$ & $0$ & $0$ & $\cdots$  & $0$ \\
	\end{tabular}
	\caption{Structure of the input data for the \EVN\ for the steady state associated to the energy profile $\Phi(\eta)=\eta$ and redshift $\kappa=0.2$; note that $1-e^{-0.2}\in\,]0.181,0.182[$. 
	}
	\label{table:input}
\end{table}


Finally, the outputs from the dense bridge and U-Net path are concatenated, and the architecture concludes with several additional dense layers to combine the information from both paths.
Overall, the network has $263\,937$ trainable parameters.
The network outputs a number $p\in\,]0,1[$. If $p\leq 0.5$, the network predicts the input steady state as unstable, otherwise it predicts the steady state to be stable.

Throughout all layers, we use $L^2$-regularization to prevent overfitting to the training data. We always use the ReLU activation function apart from the last layer where a sigmoid function is employed for binary classification. 
In order to arrive at the final architecture of the \EVN, we have implemented several stages of hyperparameter-tuning by, e.g., changing the number of filters in the convolutional layers, the number of neurons in the dense layers, the regularization parameters, etc. 

\subsection{Training}\label{ssc:traindata}

For the training of the \EVN, we have first developed an algorithm to randomly generate steady states of the Einstein-Vlasov system satisfying the conditions from Section~\ref{ssc:stst}.
From a large set of steady states generated this way, we then select interesting ones, label them, and use them as training data.
We describe these steps in more detail in the following subsections.

\subsubsection{Random steady states}\label{sssc:randomstst}

The key step to create a random steady state as in Section~\ref{ssc:stst} is to randomly choose an energy profile function~$\Phi$ satisfying~\ref{it:assphi1}--\ref{it:assphi2}.
Let us describe the process we use for this:
\begin{enumerate}[label=(\roman*)]
	\item\label{it:random1} Choose random numbers $k \in [\frac 14, 2]$ and $\eta_0 \in \,]0, 1]$.
	\item Choose a random integer $N \in \{0,\ldots,50\}$ which determines from how many parts the energy profile is generated.
	\item Generate the energy profile iteratively by adding up piecewise linear functions: We start with
	\[ 
		\Phi_0(\eta) \coloneqq \begin{cases}
			\eta, &\quad 0\leq\eta \leq \eta_0,\\
			0, &\quad \text{else}. 
		\end{cases}
	\]
	For each $i\in \{1,\ldots,N\}$, we choose random numbers $s\in [0,10]$, $l\in[\eta_0,1]$, and $r \in [l,1]$ corresponding to the incline, the left boundary, and the right boundary of the piecewise function, respectively. We then define
	\[ 
		\Phi_i(\eta) \coloneqq \begin{cases}
			s(\eta-l), &\quad  l \leq \eta \leq r,\\
			s(r-l), & \quad \eta>r,\\
			0, &\quad \text{else}.
		\end{cases}
	\]
	\item\label{it:random4}
	The final energy profile is given by 
	\[ 
		\Phi(\eta) \coloneqq \left( \sum_{i=0}^{N} \Phi_i(\eta) \right)^k,\qquad\eta\in[0,1[.
	\]
\end{enumerate}
\begin{figure}[t]
	\begin{center}
		\centering
		\input{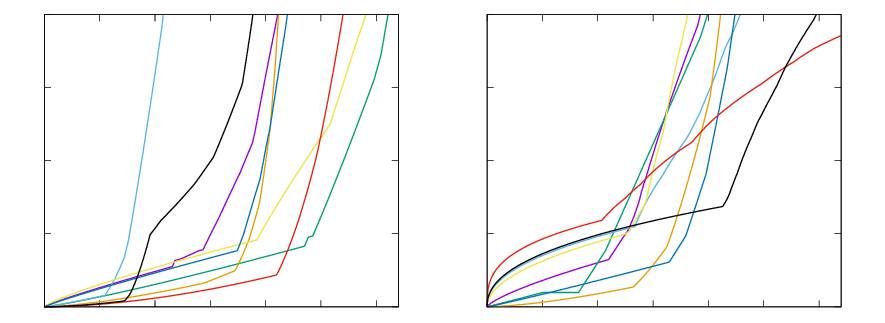}
	\end{center}
	\caption{An illustration of sixteen randomly generated energy profile functions following the process described in \ref{it:random1}--\ref{it:random4}.}
	\label{img:random_ansaetze}
\end{figure}

Finally, we choose a random redshift $0.005<\kappa<1$ where the lower limit is to prevent numerically expensive outliers; we never observe unstable steady states for such small values of $\kappa$ anyways. 

A pair of the energy profile function and the redshift~$(\Phi,\kappa)$ can be labeled as stable ($1$) or unstable ($0$) using the particle-in-cell method described in Section~\ref{ssc:numeric}.

\subsubsection{Training process}\label{sssc:train}

\begin{figure}[t]
	\centering
	\includegraphics[width=1\textwidth]{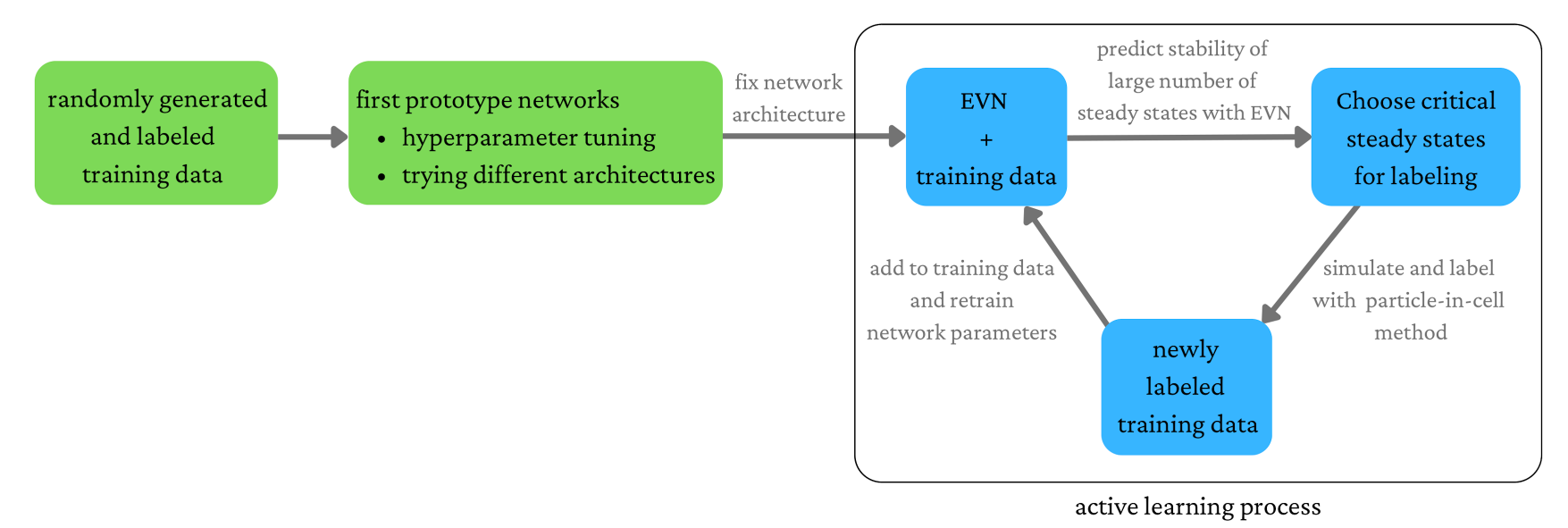}
	\caption{The training process for the \EVN\ using active learning.}
	\label{img:active_learning}
\end{figure}

The training process is illustrated in Figure~\ref{img:active_learning}.
The first part of the training process consisted of gathering a large basis of $\sim\!\!3000$ randomly generated, labeled data, according to Section~\ref{sssc:randomstst}. From this, we have implemented various versions of the final network in order to optimize accuracy, reducing variance and bias, training duration, and for fixing the hyperparameters suitably.
Throughout the entire training process, we have split all labeled data into the actual training set ($80\%$) and a cross-validation set ($20\%$). The latter was used for hyperparameter tunig and to monitor possible overfitting to the training data. 

From this point onwards, the architecture of \EVN\ is fixed and only the parameters are subject to further learning through an active learning exploration; see~\cite{KuGu20,Se09} for more background on active learning. 
Concretely, we generate a large number of random, unlabeled steady states, as pairs of $(\Phi,\kappa)$, and predict stability of the corresponding steady states via the current version of the \EVN. We then choose those pairs for further labeling which are in some sense critical cases according to one of the following criteria: 
\begin{enumerate}[label=(\roman*)]
	\item\label{it:ALELC} The predicted value $p$ is close to~$\frac 12$, i.e., the \EVN\ is unsure about the stability behavior for the particular pair $(\Phi,\kappa)$.
	\item\label{it:ALEvary} The predicted values $p_\kappa$ corresponding to $(\Phi,\kappa)$ probed along $\kappa$ vary largely in a small range of values of $\kappa$. 
\end{enumerate}
The first case~\ref{it:ALELC} is commonly known as the {\em least confidence criterion}, corresponding to data where the network is unsure or inaccurate in its prediction of the stability behavior. 
The second case~\ref{it:ALEvary} is more adapted to the present problem and detects steady states which are in some sense close to the boundary where stability changes. 
Choosing to label these critical examples via the particle-in-cell method should yield much more valuable information compared to adding more randomly picked, labeled data. After adding a suitable amount of new training data through this active learning process, we retrain the \EVN\ parameters starting from the previous parameters of the network. For the actual training process, we train in parallel multiple networks employing different number of epochs as well as various learning rate schedules. Consequently, we choose the resulting network with the lowest cross-validation error as the next version of the \EVN. 
This whole process was repeated several times. 
Throughout the training process, the cross-validation error decreased, although we added edge case steady states to the training and cross-validation sets iteratively.

The training set's final size is $8\,163$, the cross-validation set's $2\,145$. 
The network's final version achieves a cross-validation accuracy of $96.83\%$ and a training accuracy of $99.00\%$.
This difference of over two percentage points can be justified by the recurrent retraining involved in the active learning approach, which is rather sensitive to overfitting.
Nonetheless, we shall see in the following section that overfitting is not an issue on test data.


\subsection{Performance}\label{ssc:performance}

We now assess the \EVN's accuracy on three different test sets.
Before doing so, let us briefly comment on its computational costs. The \EVN\ can predict the stability of $10^3$ steady states within a few seconds when run on an ordinary laptop. 
Compared to the particle-in-cell program, which takes up to a day to determine the stability of a single steady state when run on a $40$-core supercomputer, this is rather fast.

\textbf{Performance on random test data:}
The first test set consists of $500$ randomly generated, labeled data; recall Section~\ref{sssc:randomstst} for the generation of random steady states.
On this test set, the \EVN\ achieves an accuracy of $99.00\%$, making incorrect predictions in only five cases. 

It should be noted that this error is comparable to the expected error rate resulting from numerical inaccuracies in the particle-in-cell scheme described in Section~\ref{ssc:numeric}.
Since the latter is used to label the data, we conclude that the performance of the \EVN\ on this test set is as accurate as can be reasonably hoped for.


\textbf{Performance on polytropes:}
An important class of steady states is obtained by the polytropes where the energy profile function is given by $\Phi(\eta) = \eta^k$ for $\eta>0$ and some suitable $k>0$. In the literature, it has been confirmed repeatedly~\cite{AnRe2006, Praktikum20, GueStRe21} that---for a fixed polytropic exponent~$k$---stability along the redshift $\kappa$ changes at the first local maximum of the binding energy~\eqref{eq:Eb}.
We have compared this criterion with the predictions made by the \EVN\ for the polytropic energy profiles with $\frac 14 \leq k \leq 2$ and $0.005\leq \kappa \leq 1$. The result shows agreement in $99.06\%$ of the cases. 
The errors are due to edge cases close to the point where stability changes along $\kappa$. 

We note that, in contrast to the training set and the first test set, the polytropes are not derived from the random steady state generation from Section~\ref{sssc:randomstst}.
Despite the differing data distribution of the polytropes, the \EVN's accuracy is similar as on the training set and the first test set.
This shows that the \EVN\ is not overfitted to the distribution chosen in Section~\ref{sssc:randomstst}.

\begin{figure}[h]
	\centering
	\includegraphics[width=1\textwidth]{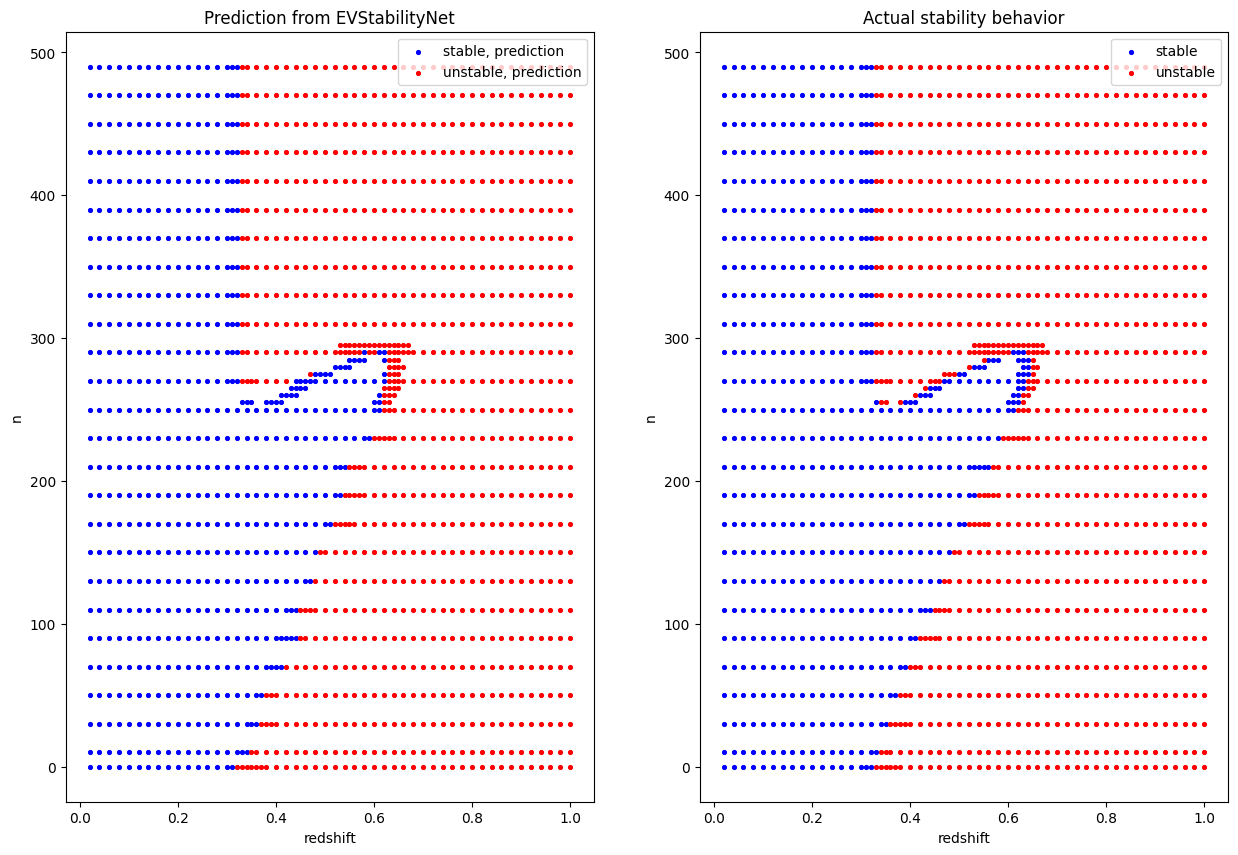}
	\caption{Performance of the \EVN\ on data from~\cite[Fig.~4]{GueStRe21}.}
	\label{img:aal_data}
\end{figure}

\textbf{Performance on data from~\cite{GueStRe21}:}
In~\cite{GueStRe21}, a class of piecewise energy profiles has been introduced  which boasts unique stability features, i.e., energy profiles which contradict the so-called strong binding energy hypothesis and ones that comprise multiple stability changes along the redshift $\kappa$. On the same set of data as illustrated in~\cite[Fig.~4]{GueStRe21}, our network correctly predicts the stability behavior in $97.43\%$ of the cases. 
The error here is larger than the test set error due to the focus on edge cases in~\cite[Fig.~4]{GueStRe21} which are inherently hard to predict correctly.  
The stability predictions of the \EVN\ on this data set alongside the actual stability behavior is depicted in Figure~\ref{img:aal_data}. 
It is remarkable that the \EVN\ can reproduce these unique stability behavior results with good accuracy, as a large amount of computational resources went into finding the results published in~\cite{GueStRe21}. 

\subsection{How to use the neural network}\label{ssc:howtouse}

The \EVN\ is available in the repository 
\begin{center}
	\href{https://github.com/Sebastian-D-G/EVStabilityNet}{https://github.com/Sebastian-D-G/EVStabilityNet}
\end{center}
for public use.
The network has been implemented and trained in Python~3.7.7 using Tensorflow~2.11.0. 
Besides the trained network, we also provide a simple working example of the network as well as the test set from above consisting of $500$ random steady states. 

We have described earlier, in particular in Section~\ref{ssc:architect}, the input that can be used for the \EVN. Let us summarize this here once more.
\begin{itemize}
	\item[(i)] The energy profile function $\Phi$ must satisfy~\ref{it:assphi1} and~\ref{it:assphi2}. It should have neither discontinuities nor too large slopes. 
	\item[(ii)] The redshift~$\kappa$ must satisfy $0.005\leq \kappa \leq 1$. 
	\item[(iii)] The input data must be of the format described in Section~\ref{ssc:architect}, see Table~\ref{table:input}.
\end{itemize}

\section{First Applications}\label{sc:applications}

The reason we developed the \EVN\ is that we can investigate the (predicted) stability of a large number of steady state with little computational costs.
Let us discuss some applications relying on this feature.


\textbf{The weak binding energy hypothesis:}
A long standing hypothesis, which has been confirmed over and over again numerically~\cite{AnRe2006, Praktikum20, GueStRe21, Ip1969_2, RaShTe1989_1,  ShTe1985_1, ShTe1985_2} and made plausible by physical reasoning~\cite{Ip1980}, is the weak binding energy hypothesis. It claims that steady states with the same energy profile function parameterized by the redshift~$\kappa$ are stable at least up to the first local maximum of the binding energy~\eqref{eq:Eb} as a function in~$\kappa$. We test the weak binding energy hypothesis by generating $2000$ random energy profiles as in Section~\ref{ssc:traindata}, determining the location of the first local binding energy maximum, and consequently predicting stability for the corresponding family of steady states with the \EVN. The results show that the weak binding energy hypothesis holds for $1984$ families of steady states. In $16$ cases the \EVN\ predicts instability closely before the first local maximum of the binding energy. However, we have checked these handful of steady states manually via the particle-in-cell method, and ascertained that these steady states are in fact stable, i.e., the prediction of the \EVN\ was slightly incorrect; this is to be expected with the accuracy of the \EVN\ at roughly $99\%$. 
Overall, we take our results as strong evidence for the validity of the weak binding energy hypothesis, as this hypothesis was never tested on such a large dataset before.

\textbf{The (strong) binding energy criterion:}
The first local maximum of the binding energy in $\kappa$ along a family of steady states often coincides with the onset of instability. For example, the polytropes seem to follow this rule, recall Section~\ref{ssc:performance}. With the same techniques as with the weak binding energy hypothesis, we have checked how often along a family of steady states, this \enquote{(strong) binding energy criterion} holds. We find that it is satisfied for $1220$ of the $2000$ energy profile functions, i.e., in $61.0\%$  of the cases. 
Given that it was shown in~\cite{GueStRe21} that the binding energy criterion does not hold in general, it is remarkable that it is nonetheless valid for such a large number of energy profiles.
Further research is required to grasp the relation between the binding energy and stability analytically.

\textbf{Negative binding energy:}
In~\cite{AnRe2006,ShTe1985_2}, it was observed numerically and discussed formally that suitable perturbations of steady states with negative binding energy~$E_b$ always result in complete dispersion. 
This was put into perspective in~\cite{Praktikum20}, where it was found that steady states with negative binding energy, which do not completely disperse, exist.
Nonetheless, it is still commonly believed that every steady state with $E_b<0$ is unstable.
We conducted an analysis of $\sim\!\!188\,000$ randomly generated steady states with negative binding energy to check the validity of this hypothesis.
The \EVN\ predicts that all of these steady states are indeed unstable, thereby further strengthening the hypothesis.

\textbf{Damping vs.\ oscillation for stable steady states:}
A future application, which we only sketch here, is to further investigate the dynamical behavior of slightly perturbed stable steady states.
In a related context, it has recently be shown that, depending on the underlying steady state, such solutions either oscillate around the original equilibrium or converge towards it in a suitable sense (\enquote{damping})~\cite{HaReScSt23}, see also~\cite{HaReSt21,RiSa2020}.
Similar behaviors have also been observed numerically for the Einstein-Vlasov system~\cite{AnRe2006,Praktikum20,GueStRe21, Gue23}. First steps towards proving the existence of oscillations around certain stable steady states in this context are made in~\cite[Ch.~6]{Gue23}.
Nevertheless, the connection between stable steady states and the presence of damped or oscillating behavior (as well as the frequency of the oscillation) is elusive for the Einstein-Vlasov system.
To advance further research, a detailed numerical study of the behavior of many slightly perturbed stable steady state will certainly be helpful.
Since one has to know which steady states are stable in the first place, the fast stability predictions of the \EVN\ can be used to significantly reduce the computational costs of such investigation.

We hope that the \EVN\ can be used for a larger range of applications beyond our initial scope, making it a valuable tool for researchers exploring the diverse field of stability of steady states of the Einstein-Vlasov system.

\end{document}